# A Month in the Life of Groupon

John W. Byers[*]
Computer Science Dept.
Boston University

Michael Mitzenmacher[†]
School of Eng. Appl. Sci.
Harvard University

Michalis Potamias[‡]
Computer Science Dept.
Boston University

Georgios Zervas[*]
Computer Science Dept.
Boston University

## ABSTRACT

Groupon has become the latest Internet sensation, providing daily deals to customers in the form of discount offers for restaurants, ticketed events, appliances, services, and other items. We undertake a study of the economics of daily deals on the web, based on a dataset we compiled by monitoring Groupon over several weeks. We use our dataset to characterize Groupon deal purchases, and to glean insights about Groupon's operational strategy. Our focus is on purchase incentives. For the primary purchase incentive, price, our regression model indicates that demand for coupons is relatively inelastic, allowing room for price-based revenue optimization. More interestingly, mining our dataset, we find evidence that Groupon customers are sensitive to other, "soft", incentives, e.g., deal scheduling and duration, deal featuring, and limited inventory. Our analysis points to the importance of considering incentives other than price in optimizing deal sites and similar systems.

## 1. INTRODUCTION

Groupon is a website offering various deals-of-the-day, with localized deals for major geographic markets. It has been one of the fastest growing Internet sales businesses in history, with tens of millions of registered users and 2011 sales expected to exceed 1 billion dollars.

We briefly describe how Groupon works; more relevant details will be given when we describe our dataset. In each geographic market, or city, there are one or more deals of the day. Generally, one deal is the featured deal of the day, and receives the prominent position on the local Groupon web site. The deal provides a coupon for some product or service at a substantial discount (generally 40-60%) to the list price. Deals may be available for one or more days. We use the term *size* of a deal to represent the number of coupons sold, and the term *revenue* of a deal to represent the number of coupons multiplied by the price per coupon. Groupon reportedly retains approximately half the revenue from the discounted coupons, and provides the rest to the seller. Groupon deals each have a minimum threshold size that must be reached for the deal to take hold, and sellers may also set a maximum threshold size to limit the number of coupons sold.

Groupon represents a change from recent Internet advertising trends. While large-scale e-mail distributions for sale offers are commonplace (generally in the form of spam) and coupon sites have long existed on the Internet, Groupon has achieved notable success with its emphasis on higher quality, localized deals, as well as its marketing savvy on both sides of this two-sided market. Understanding its success therefore seems worthwhile, and this paper represents a first step.

In this paper, we focus on how what we call "soft incentives" affect the size and revenue of Groupon deals empirically. To explain what we mean by soft incentives, we note that, obviously, the primary incentive for customers to buy is the discount Groupon offers, and naturally we examine the effect of this discount. However, Groupon also ideally tries to optimize various other parameters in managing the deals available, including how aggressively the deal is advertised, the length of time the deal is available, the days of the week it is available, and whether the deal is capped to a maximum number of purchasers. These features of deals can potentially greatly impact customer response, and hence we refer to these parameter settings collectively as soft incentives. More generally, we believe that the underlying optimization of soft incentives, such as scheduling the timing and duration of offers in a resource-constrained environment, lies at the heart of several problems in network economics, and the study of Groupon data provides an interesting setting for examining the issue in a real-world environment.

This preliminary report should be viewed as a piece of a much larger project regarding this dataset. In a forthcoming extended paper, we consider additional aspects of the data, including the predictability of deal sizes and the social processes that may affect them. The dataset will be made publicly available for other researchers (after we complete the extended paper.)

**Related Work:** To date, there has been little previous work examining Groupon specifically. Edelman et al. consider the benefits and drawbacks of using Groupon from the side of the merchant, modeling whether the advertising and price discrimination effects can make such discounts profitable [10]. Dholakia polls businesses to determine their experience of providing a deal with Groupon [9], and Arabshahi examines their business model [5]. Several works have studied other online group buying schemes that arose before Groupon, and that utilize substantially different dynamic pricing schemes [4, 12].

---

[*]E-mail: {byers, zg}@cs.bu.edu. Supported in part by Adverplex, Inc. and by NSF grant CNS-1040800.

[†]E-mail: michaelm@eecs.harvard.edu. Supported in part by NSF grants CCF-0915922, CNS-0721491, and CCF-0634923, and in part by research grants from Yahoo!, Google, and Cisco Systems.

[‡]E-mail: mp@cs.bu.edu. Supported in part by NSF grant IIS-0812309.

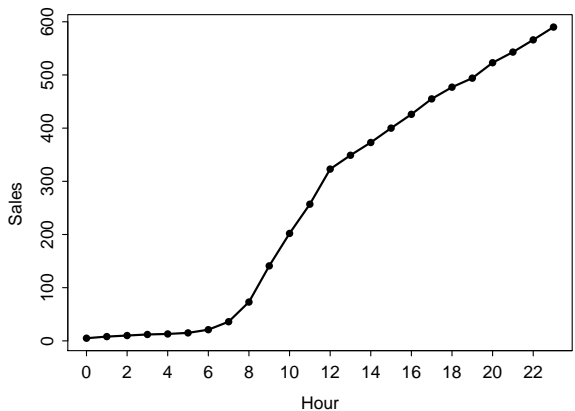

Figure 1: Cumulative hourly sales for an example deal: $10 for $20 Worth of Authentic Indian Cuisine and Drinks at Rasoi Indian Kitchen in Washington DC, Mon., 2/21/2011.

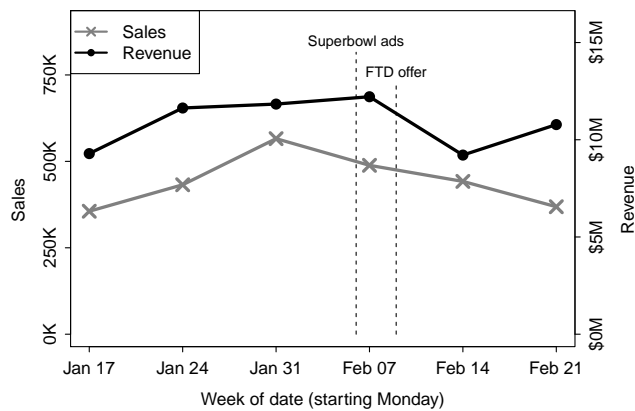

Figure 2: Weekly Groupon revenue and sales (2011).

## 2. DATASET

We collected data from Groupon between January 12th and March 3rd, 2011.[1] We monitored – to the best of our knowledge – all deals offered in 20 different cities during this period. Our criteria for city selection were population and geographic distribution. Specifically, our list of cities includes: Atlanta, Boston, Chicago, Dallas, Detroit, Houston, Las Vegas, Los Angeles, Miami, New Orleans, New York, Orlando, Philadelphia, San Diego, San Francisco, San Jose, Seattle, Tallahassee, Vancouver, and Washington DC. In total our data set contains statistics for 3729 deals.

Each deal is associated with a basic set of features: the deal description, the retail and discounted prices, the start and end dates, the threshold number of sales required for the deal to be "on", the number of coupons sold, whether the deal was available in limited quantities, and if it sold out. From these basic features we compute further quantities of interest such as the revenue derived by each deal, the deal duration, the percentage discount, and so on.

Every deal is also associated with a time-series which monitors two parameters. First, we monitor deal size growth over time; a representative time-series for a deal is shown in Figure 1. Second, we are interested on whether a particular deal was "featured" among other deals running concurrently in the same city. Featured deals are presented prominently in the webpage associated with the associated city. For example, visiting `groupon.com/boston`, one notices that a single deal occupies a significant proportion of the screen real-estate, while the rest of the deals which are concurrently active are summarized in a smaller sidebar. The effects of featuring a deal are further amplified by the daily email Groupon sends to its subscribers directs them to the webpages associated with their corresponding locations. To compile this time-series we monitored each deal in (roughly) ten-minute intervals. Occasionally some of our requests failed and therefore some gaps are present in our time-series. As the underlying processes are relatively slow-moving (for example, deals are almost always featured for at least one whole day) these gaps have minimal effect on our analysis.

Finally, we associated each deal with a category. This is an attribute which we heuristically derived by matching the description of the each deal against a set of keywords. For example, deals whose description matched keywords such as "restaurant," "food," "pizza," etc, were associated with the "Food & Drink" category. In total we identified eight broad categories: "Food & Drink", "Hotels", "Apparel", "Home Appliances", "Health & Fitness", "Tickets", "Classes", and "Autos". Deals that remained unmatched by our heuristic were placed in a special "Other" category.

We remark that our monitoring period overlapped with the immensely popular nationwide $10 for $20 Barnes & Noble gift-card offer which ran from Feb. 4th to Feb. 8th, 2011. In Los Angeles alone, this deal sold over 21,000 coupons, while the Los Angeles average deal size was approximately 622. We chose to remove this outlier deal from our dataset (with the exception of Figure 2.)

### 2.1 Insights

Figure 2 serves as an overview of the insights we are able to gain using our dataset. It displays the weekly revenue (in millions of dollars), as well as the weekly sales of coupons (in thousands) across all 20 cities we monitored. Notably, one can see the signs of fast growth associated with Groupon up to and including the week of January 31. Both sales and revenue follow an upward trend during this period. Performance appears quite different in February, where we observe an abrupt decline in both revenues and sales. February was marked by two events that generated significant negative publicity for Groupon. On February 6th Groupon's Super Bowl commercials aired, and were found offensive by many people; Groupon apologized on their official blog and pulled the controversial ad campaign [1]. Subsequently, from February 9th to February 11th, just in time for Valentine's day, Groupon offered a nationwide "$20 for $40 Worth of Flowers" coupon in partnership with FTD Flowers. Shortly thereafter, customers started realizing that when visiting the FTD Flowers website via Groupon they were shown much higher prices for the same products compared to customers who visited the FTD site directly. Furthermore, coupon purchasers were only allowed to use their coupons on the more expensive version of the product. The so called offer was variously perceived as anywhere between a "bait & switch" scheme and an outright scam; Groupon offered disgruntled customers refunds[2] [3]. The recovery that we observe in the revenue indicates that this shock had a short-term effect on Groupon's growth. Similar short-term effects have been observed in the time-series of YouTube video views after exogenous shocks [8].

---

[1] We continue to collect data. The process is made more difficult as Groupon from time to time changes not only its page format, but also the URLs required to access information on deals. We expect our final release of data will contain well over 10,000 deals.

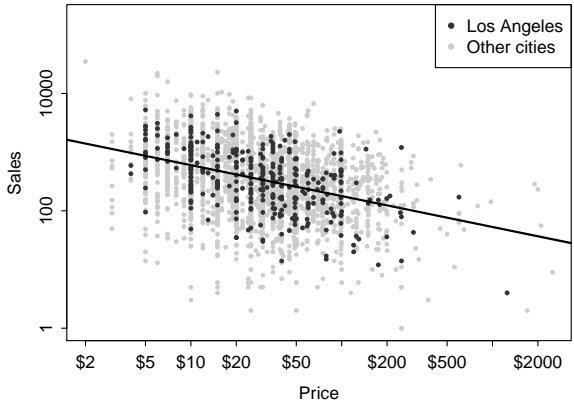

Figure 3: Deal size by price (in logarithmic scale.) Black dots highlight deals in LA; grey dots are used for other cities. A trend line fitted using OLS regression is also shown.

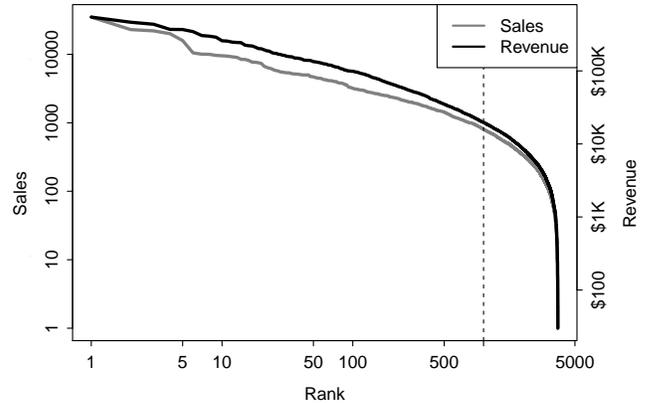

Figure 4: Groupon sales and revenue by rank.

Entertaining anecdotes aside, using our dataset we are able to glean various operational insights on and performance metrics for Groupon's business. Figure 4 displays all deals in our dataset ranked by size and revenue. The axes are in logarithmic scale. Both deal sizes and deal revenues follow a power-law-like distribution for much of the distribution, but the distribution changes for small deals that earned less than $1K of revenue or had fewer than 100 sales. Looking at the top 1000 deals we find that 46% of them were featured, compared to just 17% for the remaining deals. The mix of other deal attributes is similar with one exception: 46% of the top 1000 deals come from the "Food & Drink" category, whereas the corresponding percentage for remaining deals is only 21%. In future work we plan to examine this distribution further; we suspect that different social processes may be affecting the small and large deals.

## 3. DEAL FEATURES AND INCENTIVES

Groupon has built its business around offering deep discounts. As such one would expect that the discount associated with a deal would be the primary purchase incentive. Figure 3 is a scatterplot of deal sizes (coupons sold) vs. deal prices in logarithmic scale. We observe a large amount of variance within individual price points. By controlling for other features we can gain a clearer picture of the trend, as can be seen by the highlighted data for Los Angeles. Fitting a trend line on the full data set using ordinary least-squares (OLS) regression provides more perspective on the dependence between price and deal size. While price alone is not enough to predict deal size, it appears that the logarithm of price and logarithm of sales are roughly linearly related.

One feature is closely related to the price of a deal: its list price. The two are displayed side by side. In fact, Groupon has developed a set of editorial guidelines [2]; deal titles commonly take on a standard form: "$X for $Y Worth of Z" (eg, "$10 for $20 Worth of Books at Books Inc.")  As such one might think that presenting the list price next to the discounted price might signal how good a bargain a certain deal is and provide a further purchase incentive to customers. However, in our dataset approximately three quarters of all deals are discounted by 40 to 60%. The list price is highly correlated (0.87) with the discount price and therefore a weak sales-distinguishing feature once price has been taken into account.

Approximately 46% of all Groupon deals in our dataset were scheduled to run for 24 hours. About 32% ran for two days, 18% for three days, and the remaining few for four days or more. Figure 5c indicates that while deal size hardly varies across deals with different durations, revenue increases for deals that run longer. This may suggest that Groupon is attempting to hit *sales*, instead of *revenue*, targets. Deals that are more expensive must be allowed more time to achieve the same sales goal. When we break down deals by their duration and compute their mean prices we observe that longer deals are, on average, more expensive:

| Duration (days) | 1 | 2 | 3 | 4+ |
|---|---|---|---|---|
| Mean price | $27 | $41 | $67 | $148 |
| Number of deals | 1,731 | 1,193 | 680 | 125 |

We observe from the distribution of daily sales for multi-day deals that for two-day deals approximately 65% of total sales occurred on the first day, while for three-day deals 56% of total sales occurred on the first day, 22% on the second, and the remaining on the final day.

At any point in time, and for each geographic market,

---
[2]Since we cannot account for these refunds, we note that the reported sales and revenues of the week starting on February 7 are an overestimate.

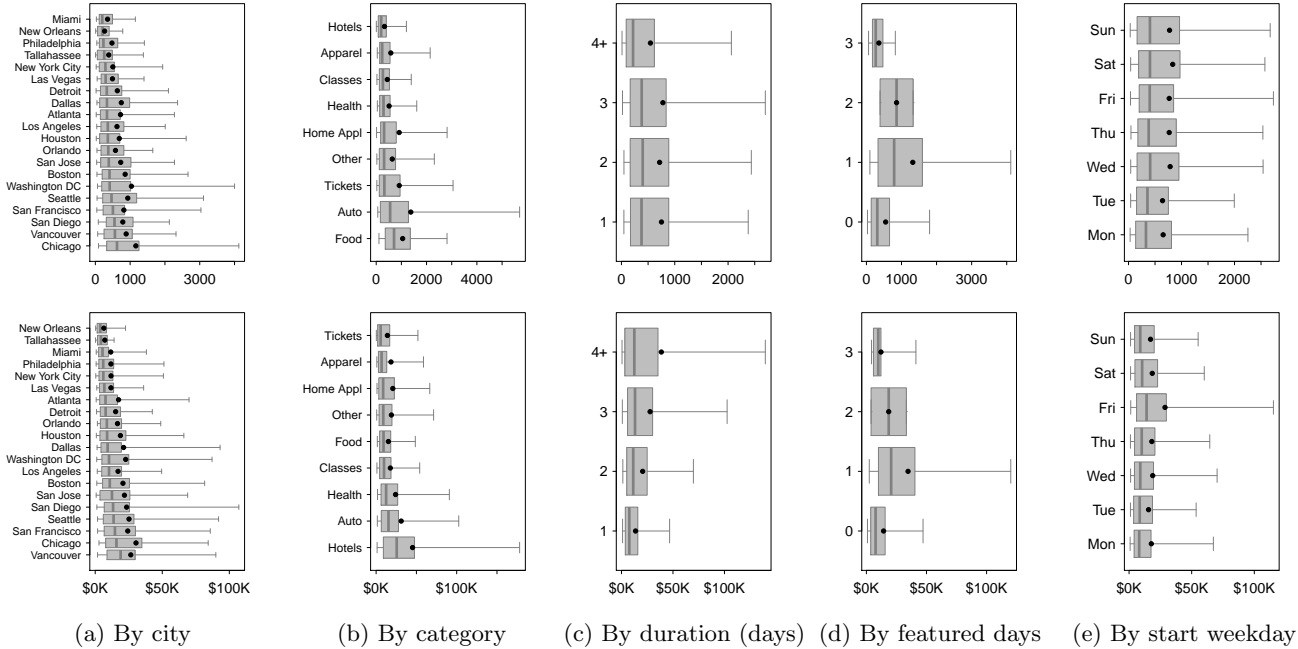

(a) By city  (b) By category  (c) By duration (days)  (d) By featured days  (e) By start weekday

Figure 5: Deal sizes, on the top row, and deal revenues, on the bottom row are shown across the $x$−axis. On the $y$−axis deals are broken down by various deal features. The outer whiskers mark the 5% and 95% quantiles of the deal size and revenue, the sides of each grey rectangle the 1st and 3rd quartiles, the solid bar the median, and the solid dot the mean.

one deal among the set of all available deals is featured by Groupon. Featured deals receive prominent placement both on the Groupon website and the daily email that customers receive. In our dataset, 75% of all deals were never featured, 24.5% were featured for a single day, and 0.5% were featured for 2 or 3 days. For simplicity, we therefore treat featuring as a boolean attribute: a deal was either featured, or not. The impact of featuring deals is significant: the mean sales and revenue for featured deals comfortably exceed twice the corresponding quantities for other deals. The effects of featuring a deal are summarized in the following table:

|  | Featured | Non-featured |
| ---: | ---: | ---: |
| Mean size | 1,310 | 548 |
| Mean revenue | $34,354 | $14,277 |
| Number of deals | 2,799 | 930 |

However, we cannot assume that the effect of featuring a deal is entirely causal. Groupon has an incentive to feature those deals that it projects will be most attractive to their customers. A natural question which follows is whether certain categories of deals might as whole be more attractive to customers. Table 1 breaks down deal categories by the percentage of deals within each that was featured. With the exception of "Apparel" and "Tickets" deals, there is little variance across categories in terms of featuring them.

Another distinguishing characteristic of a deal is the inventory size: some deals are available only in limited quantities, while others have unlimited inventory. Groupon lets its customers know which deals are limited, but does not display the number of available units (while some competitors of Groupon display this prominently). Approximately 29% of all deals in our dataset are available in limited numbers, with wide variation across categories. Approximately 44% of all "Tickets" deals and 77% of all "Hotels" deals were avail-able in limited quantities, as one might expect for these types of deals. Surprisingly, only about 18% of the limited deals in our dataset sold out. It is not that these deals are intrinsically less attractive: limited deals outperformed unlimited deals on average by 8% more coupon sales and achieving 20% more revenue. Perhaps sellers or Groupon artificially "limit" deals to apply pressure to customers, making them more likely to purchase on the spur of the moment.

|  | Featured | Non-featured | # of deals |
| ---: | ---: | ---: | ---: |
| Apparel | 86% | 14% | 125 |
| Auto | 72% | 28% | 71 |
| Classes | 76% | 24% | 352 |
| Food & Drink | 73% | 27% | 1,021 |
| Health & Fitness | 78% | 22% | 761 |
| Home Appliances | 74% | 26% | 91 |
| Hotels | 77% | 23% | 70 |
| Other | 76% | 24% | 994 |
| Tickets | 66% | 34% | 244 |

Table 1: A breakdown of featured deals by category.

Groupon also has a choice as to the day(s) of the week that it schedules each deal to run. Figure 5e breaks down deals by the day of the week on which they began. Even though the differences are not striking, it appears deals starting in the beginning of the week produce less revenue, and deals starting on Friday produce the most. One possible explanation is that on Fridays Groupon starts more multi-day deals that span the weekend. For example, 61% of three-day deals start on a Friday. However, alternative explanations, such as that the best deals are run on Friday or that people are more likely to buy on Friday, may also apply.

Deals are said to be "on" when they surpass a sales thresh-old defined by Groupon, possibly in conjunction with the merchant. That is, for customers to get a discount, a minimum number of them must commit to the deal. Theoretically this could yield a group dynamic where customers encourage their friends to buy a coupon to make it "on." However, current deal thresholds are very low. For example, in our dataset, the mean threshold to total sales ratio is approximately 20%. Deals that surpassed their threshold did so on average just after 8am, and only 2% of deals with thresholds failed to reach them. Like the relationship between price to deal size, the relationship between threshold and deal size also appears to be linear when plotted in logarithmic scale (plot omitted for space).

Deals can also be classified by two features that Groupon does not have direct control over: their geographic market, and their category. Figure 5a shows that overall deal sizes appear quite similar over all major cities. One might expect city population to have a more pronounced effect. However, New York does not include Long Island, which has a separate deal stream; similarly Los Angeles is split into multiple subareas. As far as deal categories are concerned, Figure 5b suggests that deals that are the most lucrative in terms of revenue for Groupon are not the most popular with their customers. The most striking example is "Hotels" deals. Even though on average they produce the fewest sales, they are also the most profitable. "Food & Drinks" deals prove to be the most popular, but produce far less revenue.

Finally, it is interesting to note that throughout Figure 5, we observe heavy-tailed behavior, with the mean being much larger than the median in essentially all cases.

## 4. MODELING DEAL OUTCOMES

In the previous section, we considered the soft incentives available to Groupon (as well as other features), and their individual relationships with deal size and revenue. We proceed to consider these characteristics in a unified manner by considering a regression model. The purpose of this is twofold: first, it allows us to quantify the dependence of deal outcomes on the various features we have described; second, if sufficiently accurate, we could use the model to predict the outcomes of future deals. (Prediction will be a major topic in our later paper.) The model we use, noting that all logarithms are base $e$, is:

$$\log q = \beta_0 + \beta_1 \log p + \beta_2 \log t + \beta_3 d + \beta_4 f + \beta_5 l \\ + \bar{\beta}_6 \mathbf{w} + \bar{\beta}_7 \mathbf{c} + \bar{\beta}_8 \mathbf{g} \quad (1)$$

where $q$ stands for the deal size, $p$ for the coupon price, $t$ for the threshold, $d$ for whether the deal run for multiple days or not, $f$ for whether the deal is featured or not, and $l$ for whether the deal inventory is limited or not. The values of $p$ and $t$ are centered to their corresponding medians (25 in both cases.) This allows for a more intuitive interpretation of the regression's intercept but does not otherwise affect our results. The parameters $\mathbf{w}$, $\mathbf{c}$ and $\mathbf{g}$ are dummy-coded vectors representing the starting day of the week, category, and city relative to notional reference levels; their corresponding coefficients are also vectors. Dummy-coding refers to using binary vectors to encode categorical variables, where a variable that can take on $k$ distinct values is encoded using a binary vector of length $k-1$ where at most one entry is set to one.

The exact form the model takes upon is motivated by the observations of Section 3: $\log p$ and $\log q$ appear to be linearly related to $\log q$, while the rest of the variables in our model are either boolean or categorical in nature. Given the high correlation of the list price to the discounted price we have chosen to exclude it from the model to avoid introducing multicollinearity. Also, since most multi-day deals last for two days, and there is little variance in the number of sales among multi-day deals, we have chosen to encode duration as a boolean feature.

We fitted the model using OLS regression. The parameter estimates, their standard errors, and their significance levels are shown in Table 2. The intercept of the model is the unconditional expected mean of the logarithm of the size of a deal. That is to say the expected mean of the log-size of a deal in the "Other" category, priced at $25, with a threshold of 25, not featured, with unlimited inventory, starting on Monday, and running for a single day in Atlanta is 5.32. Equivalently, the expected *geometric* mean of the size of the same deal is $e^{5.32} \approx 204$.

The coefficient of $\log p$ is particularly interesting because its value is the *point-price elasticity of demand* for coupons. To see this, recall that, by definition, the point-elasticity $\eta_p$ is given by:

$$\eta_p = (\partial q / q)/(\partial p / p) = \partial \log q / \partial \log p = \beta_1. \quad (2)$$

|  |  | Estimate | Std. Error | Signif. |
|---|---|---|---|---|
| Incentives | (Intercept) | 5.32 | 0.08 | *** |
|  | log Price | -0.47 | 0.02 | *** |
|  | log Threshold | 0.47 | 0.02 | *** |
|  | Multi-day? | 0.27 | 0.04 | *** |
|  | Featured? | 0.81 | 0.04 | *** |
|  | Limited? | -0.02 | 0.04 |  |
| Start Day | Mon | *Reference level* |  |  |
|  | Tue | 0.00 | 0.05 |  |
|  | Wed | 0.17 | 0.05 | ** |
|  | Thu | 0.19 | 0.05 | *** |
|  | Fri | 0.20 | 0.06 | *** |
|  | Sat | 0.03 | 0.06 |  |
|  | Sun | -0.04 | 0.06 |  |
| Category | Other | *Reference level* |  |  |
|  | Apparel | -0.08 | 0.09 |  |
|  | Autos | 0.25 | 0.12 | * |
|  | Classes | -0.16 | 0.06 | ** |
|  | Food & Drink | 0.17 | 0.05 | *** |
|  | Health & Fitness | 0.02 | 0.05 |  |
|  | Home Appliances | -0.05 | 0.10 |  |
|  | Hotels | -0.11 | 0.12 |  |
|  | Tickets | -0.29 | 0.07 | *** |
| City | Atlanta | *Reference level* |  |  |
|  | Boston | 0.21 | 0.09 | * |
|  | Chicago | 0.48 | 0.08 | *** |
|  | Dallas | -0.01 | 0.09 |  |
|  | Detroit | -0.10 | 0.10 |  |
|  | Houston | -0.26 | 0.10 | ** |
|  | Las Vegas | -0.21 | 0.10 | * |
|  | Los Angeles | 0.13 | 0.08 |  |
|  | Miami | -0.37 | 0.10 | *** |
|  | New Orleans | -0.58 | 0.13 | *** |
|  | New York | -0.18 | 0.08 | * |
|  | Orlando | -0.00 | 0.11 |  |
|  | Philadelphia | -0.32 | 0.10 | ** |
|  | San Diego | 0.05 | 0.10 |  |
|  | San Francisco | 0.38 | 0.09 | *** |
|  | San Jose | -0.02 | 0.12 |  |
|  | Seattle | 0.31 | 0.09 | *** |
|  | Tallahassee | -0.58 | 0.16 | *** |
|  | Vancouver | 0.15 | 0.11 |  |
|  | Washington DC | 0.34 | 0.09 | *** |
|  | $F$-statistic | 79.8 | $p$-value | < 2e-16 |
|  | $R$-squared | 0.45 |  |  |
|  | Significance codes | 0% *** 0.1% ** 1% * 5% |  |  |

Table 2: The regression model of deal sizes on their features.

For deal size, the point-price elasticity given by our regression model is $-0.47$. Intuitively, this means that for any 1% increase in price, we expect a (roughly[3]) 0.47% decrease in demand. Since $\eta_p > -1$ the demand is said to be inelastic. This matches our intuition: coupons already represent heavy discounts and as such neither an increase nor a decrease in price should severely affect demand.

The coefficients of non-log-transformed variables represent differences in the expected means of log-sales, and their exponentiated values represent multiplicative increases (or decreases) in the expected geometric mean of sales. For example, the expected ratio of the geometric means of sales for multi-day deals to single-day deals is $e^{0.27} \approx 1.31$. In simple terms, we could claim that running a deal for more than a day, we expect a 31% increase in sales. The effect of featuring a deal is, as anticipated, far greater: featured deals are expected to perform 127% better than their non-featured counterparts. Limiting a deal, in the presence of all the other parameters in our model, does not have a statistically significant effect.

Overall, the $F$-statistic and the $p$-value of the model indicate that we can reject the null hypothesis (that there is no relation between sales and any of the explanatory variables in our model) with high confidence. However, the $R$-squared value of the model suggests only moderate predictive power. Improving the predictive power of the model is ongoing work.

## 5. DEAL SCHEDULING

One optimization Groupon has available is managing how deals are scheduled. Here we examine the data to observe whether there are noticeable scheduling artifacts demonstrating active management by Groupon. Of course this raises the interesting question of how to best schedule deals to optimize revenue or sales; we do not consider this question here, but note that similar problems have been studied in advertising (e.g., [11, 14]).

As previously alluded to, we note that Groupon does some scheduling according to the day of the week. For example, 66% of all deals start Monday through Thursday; on Saturdays and Sundays, Groupon initiates approximately 40% and 50% fewer deals respectively than on Mondays. For another example, in Figure 6, we plot the duration of deals by the day of the week that they started on. The subplots show the counts of one-day, two-day, and three-day deals by their starting day of the week. We observe that Groupon initiates most one-day deals Monday through Thursday and most three-day deals on Fridays and weekends.

We also find evidence that deals are scheduled according to their category; specifically, that Groupon avoids placing featured deals of the same category back-to-back over consecutive days in the same city. This is a natural strategy to maintain user interest, related to issues of ad fatigue (or advertising wear out) studied in other contexts [6, 7].

Given the original time-series of featured deals for each city, for each pair of categories, we count how many times a deal of category $A$ was followed by a deal of category $B$. (As we only consider featured deals, the number of deals is a small fraction of our total dataset.) We also compute

---

[3]The point-price elasticity relates infinitesimal percentage changes in price to percentage changes in demand. For any small enough percentage change in price the estimated percentage change in sales will be reasonably accurate.

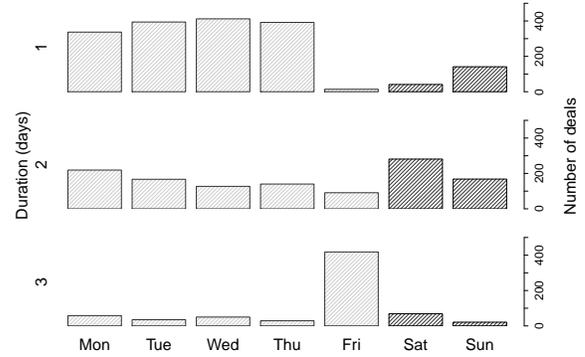

Figure 6: The number of deals starting on any given day of the week, broken down by their eventual duration.

**Following category**

|  | Ap. | Au. | Cl. | Fo. | He. | Hm. | Ht. | Ot. | Ti. |
|---|---|---|---|---|---|---|---|---|---|
| Ap. | 0/0 | 0/0 | 1/1 | 7/6 | 3/4 | 1/1 | 1/0 | 3/4 | 1/2 |
| Au. | 0/0 | 0/0 | 4/2 | 6/6 | 2/3 | 1/1 | 1/0 | 3/5 | 3/2 |
| Cl. | 0/1 | 3/2 | 4/7 | 32/26 | 14/14 | 2/2 | 1/2 | 20/21 | 8/7 |
| Fo. | 9/6 | 7/6 | 27/26 | 57/86 | 62/47 | 4/7 | 5/5 | 75/68 | 26/24 |
| He. | 6/4 | 4/3 | 15/14 | 60/48 | 17/29 | 6/4 | 4/2 | 38/43 | 14/14 |
| Hm. | 0/0 | 0/1 | 1/2 | 6/7 | 9/4 | 1/0 | 0/0 | 6/6 | 0/2 |
| Ht. | 0/0 | 2/0 | 1/2 | 6/5 | 2/3 | 1/0 | 0/0 | 3/4 | 1/1 |
| Ot. | 2/4 | 3/5 | 18/21 | 70/67 | 42/44 | 6/6 | 2/4 | 67/62 | 24/21 |
| Ti. | 1/2 | 0/2 | 11/7 | 29/24 | 12/14 | 1/2 | 2/1 | 22/21 | 2/7 |

(with "Preceding" labeling rows)

Table 3: The observed/expected number of times one featured category follows another.

the average value of this quantity by randomly permuting the time-series 1,000 times. Table 3 summarizes the experiment. For each pair of categories we show the observed number of occurrences of that event followed by the average number of occurrences (to the nearest integer). For example, the entry 29/24 of preceding category "Ti." and following category "Fo." indicates that a featured deal of category "Tickets" was followed by a featured deal of category "Food" 29 times in our data, while if the same deals had been scheduled randomly, we would expect this event to happen 24 times. Deviations from the random experiment are generally small, *except* for the diagonal of this matrix, which indicates that Groupon consistently avoids featuring the same category back-to-back. The observation holds for all categories except for the "Other" category, which contains all deals that do not fit in any of our broad categories.

## 6. CONCLUSION AND FUTURE WORK

Our work to date has characterized customer purchases of daily deals offered by Groupon, and has investigated relationships between price, other key deal attributes, and deal size. One key outcome of our modeling work is preliminary evidence that Groupon is behaving strategically to optimize deal offerings, giving customers "soft" incentives to purchase.

One specific line of our future work regards the predictability of deal sizes. How early in the day can we obtain strong predictions of the final deal size based on sales thus far and our time-series data, and how does this compare with our zero hour prediction? Another line examines the interplay between social networks and deal purchases. To what extent does sharing of information through social networks increase Groupon sales, and do existing propagation models, such as the Independent Cascade model [13], apply?